\documentclass[aps,prd,superscriptaddress,showpacs,amsmath,amssymb,nofootinbib]{revtex4}

\usepackage[final]{graphicx}
\usepackage[linktocpage]{hyperref}
\hypersetup{colorlinks=false, pdfborder={0 0 0}, pdftitle={Trinification, the Hierarchy Problem and Inverse Seesaw Neutrino Masses}}

\usepackage[american]{babel}
\usepackage[utf8]{inputenc}
\usepackage[T1]{fontenc}
\usepackage{enumerate}
\usepackage{mdwlist}
\usepackage[activate=normal]{pdfcprot}
\usepackage[babel]{csquotes}
\usepackage{bbding}
\usepackage{color}

\usepackage{mathrsfs}
\usepackage{slashed}
\usepackage{bbm}
\usepackage{feynmp}
\usepackage{wrapfig}
\usepackage{color}

\newcommand{\order}[1]{\mathcal{O}\left({#1}\right)}

\begin{document}

\begin{flushright}
DO--TH--10/19
\end{flushright}

\title{Trinification, the Hierarchy Problem and Inverse Seesaw Neutrino Masses}
\author{Christophe Cauet}
\email{christophe.cauet@tu-dortmund.de}
\author{Heinrich P\"as}
\email{heinrich.paes@tu-dortmund.de}
\affiliation{Fakult\"at f\"ur Physik, Technische Universit\"at Dortmund, 44221 Dortmund, Germany}
\author{S\"oren Wiesenfeldt}
\email{soeren.wiesenfeldt@helmholtz.de}
\affiliation{Helmholtz Association, Anna-Louisa-Karsch-Str.\,2, 10178 Berlin, Germany} 

\begin{abstract}
In minimal trinification models light neutrino masses can be generated via a radiative see-saw mechanism, where the masses of the right-handed neutrinos originate from loops involving Higgs and fermion fields at the unification scale. This mechanism is absent in models aiming at solving or ameliorating the hierarchy problem, such as low-energy supersymmetry, since the large seesaw-scale disappears. In this case, neutrino masses need to be generated via a TeV-scale mechanism. In this paper, we investigate an inverse seesaw mechanism and discuss some phenomenological consequences.
\end{abstract}

\maketitle
\section{Introduction}
Trinified models, $SU(3)_C\times SU(3)_L\times SU(3)_R$, are attractive candidates for a more symmetric extension of the Standard Model (SM) as all matter can be arranged in bi-fundamental representations; no adjoint Higgs representations are needed to break the symmetry to the Standard Model; and gauge interactions conserve baryon number, thus proton decay is naturally suppressed \cite{achiman1978,kang1985}.  Moreover, trinified models can be motivated as the low energy limit of string theory, both as a subgroup of $E_8$ in heterotic string theory \cite{gross1984} and $N=8$ supergravity \cite{cremmer1979} as well as in IIB string theories via AdS/CFT duality with a conformal $SU(3)^n$ gauge theory (see e.g.  \cite{frampton1999,kephart2001,kephart2004})\footnote{Investigations of trinified models have also been performed in the context of extra dimensions, see e.g. \cite{carone2004,carone2005,cacciapaglia2006}.}.
A generic problem arising in all theories which aim at unifying the gauge interactions at some large unification scale $M_U$ is the hierarchy problem: the large hierarchy between the unification and the electroweak scale is unstable against radiative corrections.  The most popular way out is to introduce weak-scale supersymmetry which gives rise to the cancellation of radiative corrections of SM particles and their superpartners.  The additional particle content can ensure gauge coupling unification at $M_U$; moreover, there exists also a natural dark matter candidate.  Alternatives of supersymmetry which adress at least some of these points and in particular the hierarchy problem include theories with large extra dimensions \cite{antoniadis1998,arkanihamed1998,randall1999a,randall1999b}, a large number of copies of the SM states \cite{dvali2009} or models based on AdS/CFT complementarity \cite{frampton1999}.  In such scenarios the Planck scale is typically lowered to the electroweak scale, avoiding any high energy scale in the theory.

As we will see, the extension of trinified models with either of these ideas leads to important consequences for the mechanism of neutrino mass generation, namely the absence of a large seesaw scale and the necessity to generate neutrino masses at the TeV scale.  In this paper we study the implementation of TeV neutrino mass generation via an inverse seesaw mechanism and some of its phenomenological consequences. We will show that in general, neutrino masses can be generated in the desired range.

\section{ A short review on minimal trinification}
We begin by briefly reviewing the trinified model \cite{achiman1978,kang1985,pakvasa1986,mt2006}. Gauge coupling unification is guaranteed by an additional discrete $\mathbb{Z}_3$ symmetry, which results in \enquote{minimal trinification}, $SU(3)_C \times SU(3)_L \times SU(3)_R \times \mathbb{Z}_3$. The fundamental representation of $SU(3)_C\times SU(3)_L\times SU(3)_R$ is $(1,3,3^*)\oplus(3^*,1,3)\oplus(3,3^*,1)$, which forms the fundamental fermion representation $\mathbf{27}$ of $E_6$ \cite{slansky1981}.  The fermion multiplets are assigned to the irreducible representations as follows:
\begin{align}
  \psi_L\oplus\psi_{Q^c}\oplus\psi_Q\equiv (1,3,3^*)\oplus(3^*,1,3)\oplus(3,3^*,1).
\end{align}
With respect to the SM ($G_\text{SM}=SU(3)_C\times SU(2)_L\times U(1)_Y$) the fermion multiplets decompose into:
\begin{align}
  \psi_L&\rightarrow(1,2,\tfrac{1}{2})\oplus 2(1,2,-\tfrac{1}{2})\oplus(1,1,1)\oplus 2(1,1,0), \nonumber\\
  \psi_{Q^c}&\rightarrow(3^*,1,-\tfrac{2}{3})\oplus 2(3^*,1,\tfrac{1}{3}),\\
  \psi_Q&\rightarrow(3,2^*,\tfrac{1}{6})\oplus(3,1,-\tfrac{1}{3})\nonumber
\end{align}
The Hypercharge $Y$ is given by the Gell-Mann-Nishijima formula $Y=Q+I_3$, where $I_3$ is the third component of the $SU(2)_L$ isospin and $Q$ is the electric charge.  The SM charged leptons and the neutrino of each generation are accomodated in the same $\psi_L$ multiplet,
\begin{align}
  \psi_L = \begin{pmatrix}
      	(\mathscr{E}) & (E^c) & (\mathscr{L})\\
    	\mathscr{N}_1 & e^c & \mathscr{N}_2
			\end{pmatrix}.
\end{align}
A superposition of $\mathscr{E}$ and $\mathscr{L}$ forms the known standard model weak lepton doublet whereas $e^c$ is the common positron field.  $E^c$ is a new lepton doublet with the opposite hypercharge compared to $\mathscr{E}$ and $\mathscr{L}$. $\mathscr{N}_1$ and $\mathscr{N}_2$ are heavy neutral leptons, and therefore sterile $SU(2)$ singlet neutrinos. The right-handed and left-handed quarks are embedded in $\psi_{Q^c}$ and $\psi_Q$, respectively,
\begin{align}
  \psi_{Q^c}=\begin{pmatrix} \mathscr{D}^c\\
    u^c\\
    \mathscr{B}^c
   \end{pmatrix},\qquad
  \psi_Q&=\begin{pmatrix}
    (-d & u) & B
   \end{pmatrix}.
\end{align}
Here $u^c$ is the common up-conjugate quark field whereas $\mathscr{D}^c$ and $\mathscr{B}^c$ are a superposition of the common $d^c$ quark and a new heavy quark $B^c$ carrying the same quantum numbers. The doublet $(-d\;u)$ is the conjugate of the usual quark doublet $Q=\tbinom u d$. 

When the GUT symmetry is broken, as discussed below, a linear combination of $\mathscr{D}^c$ and $\mathscr{B}^c$ together with $B$ will become massive with mass $\mathcal{O}(M_{\mathrm{GUT}})$ whereas the orthogonal combination remains light. This combination is the down-quark singlet field of the SM. We denote the mixing angle between $\mathscr{D}^c$ and $\mathscr{B}^c$ as $\alpha$ such that\footnote{There are small corrections $\mathcal{O}(M_{\mathrm{EW}}/M_{\mathrm{GUT}})$ when the electroweak symmetry is broken and the down quark acquires mass. For three generations, this relation expands to a six-by-six mixing matrix for $(d,B)$ and $(\mathscr{D}^c,\mathscr{B}^c)$, which includes the CKM matrix. For more details, see Ref. \cite{mt2006}.}
\begin{align}
	\begin{pmatrix}
          	d^c\\
		B^c
        \end{pmatrix}=
	\begin{pmatrix}
          	-s_\alpha\,\mathscr{D}^c+c_\alpha\,\mathscr{B}^c\\
		c_\alpha\,\mathscr{D}^c+s_\alpha\,\mathscr{B}^c
    \end{pmatrix}.
\end{align}
For brevity we write $c_\alpha=\cos\alpha$ and $s_\alpha=\sin\alpha$. Similarly, we denote the mixing angle in the lepton sector as $\beta$ such that
\begin{align}
	\begin{pmatrix}
        E\\
		L
        \end{pmatrix}=
	\begin{pmatrix}
          	-s_\beta\,\mathscr{E}-c_\beta\,\mathscr{L}\\
		c_\beta\,\mathscr{E}-s_\beta\,\mathscr{L}
        \end{pmatrix}, \qquad
		N_1 = s_\beta\,\mathscr{N}_1-c_\beta\,\mathscr{N}_2, 
		\qquad\text{and}\qquad
		N_2 = -c_\beta\,\mathscr{N}_1-s_\beta\,\mathscr{N}_2.
\end{align}

Two $(1,3,3^*)$ Higgs multiplets $\varphi_L^{1,2}$ are used to break down both the electroweak and the unified symmetry \cite{pakvasa1986}. The vacuum expectation values (VEVs) $v_i$ break the trinified group at the unification scale $M_{\mathrm{U}}$, while $u_i$ and $n_i$ are $\mathcal{O}(M_{\mathrm{EW}})$ and thus break the SM group,
\begin{align}
 \varphi_L^a=\begin{pmatrix}
    (\varphi_1^a) & (\varphi_2^a) & (\varphi_3^a)\\
    S_1^a & S_2^a & S_3^a
  \end{pmatrix},\quad \langle\varphi_L^1\rangle=\begin{pmatrix}
    u_1 & 0 & 0\\
    0 & u_2 & 0\\
    0 & 0 & v_1
  \end{pmatrix},\quad \langle\varphi_L^2\rangle=\begin{pmatrix}
    n_1 & 0 & n_3\\
    0 & n_2 & 0\\
    v_2 & 0 & v_3
  \end{pmatrix}.
\end{align}

Due to the quantum number assignment of the $(1,3,3^*)$ multiplet, this is the most general expression for the VEVs.  In order to generate up-type quark masses, one of $u_2$ and $n_2$ needs to be  nonzero; similarly, at least one of $u_1$, $n_1$, and $n_3$ is necessary for the down-type quark and the charged lepton masses.

Using the minimal set of VEVs required to obtain the correct fermion masses, $u_1$, $u_2$, $v_1$, and $v_2$ are chosen to be nonzero and all others zero.The discrete $\mathbbm{Z}_3$ symmetry requires colored Higgs fields $\varphi_{Q^c}^a$ and $\varphi_Q^a$ in addition to the Higgs multiplets  $\varphi_L^a$, 
\begin{align}
    \varphi_{Q^c} = \begin{pmatrix}
      \mathscr{D}_H^c\\
      \mathscr{U}_H^c\\
      \mathscr{B}_H^c
    \end{pmatrix}
    ,\qquad \varphi_Q=
    \left(-\mathscr{D}_H\;\mathscr{U}_H\;\mathscr{B}_H\right) .
  \end{align}

The general Higgs potential was studied in Ref.~\cite{pakvasa1986}.  For our purpose, however, it is sufficient to restrict ourselves to a simplified case which considers only one of the two possible Higgs multiplets, $\varphi_L\equiv\varphi_L^1$, as well as only the dimension-two and three terms.  In this case, the potential simply reads
  \begin{align}\label{eq:HiggsPotential}
    \mathscr{L}_h & = m^2 \left(\varphi_Q^*\varphi_Q +
      \varphi_{Q^c}^*\varphi_{Q^c} + \varphi_L^*\varphi_L \right)
    + \left[ \gamma_1\varphi_{Q^c}\varphi_Q\varphi_L + \gamma_2
      \left(\varphi_L\varphi_L\varphi_L+\text{cyclic} \right) +
      \text{h.c.}  \right],
  \end{align}

where $m,\gamma_i=\mathcal{O}(M_U)$.  The bilinear terms are mass terms proportional to $m^2$, while the cubic terms are proportional to $\gamma_i$.

The Yukawa couplings are given by two types of interactions, which are allowed due to their singlet structure under gauge group transformations,
\begin{align}
  \psi_{Q^c}\psi_Q\varphi_L^a & \equiv
  (\psi_{Q^c})^i_j(\psi_Q)^j_k(\varphi_L^a)^k_i,\nonumber  \\
  \psi_L\psi_L\varphi_L^a & \equiv \varepsilon^{ijk} \varepsilon_{rst}
  (\psi_L)^r_i (\psi_L)^s_j (\varphi_L^a)^t_k .
\end{align} 
and their cyclic permutations. The general Yukawa couplings for Quarks and Leptons are therefore
\begin{align}
  \mathscr{L}_q & = g \left( \psi_{Q^c}\psi_Q\varphi_L +
    \psi_L\psi_{Q^c}\varphi_Q + \psi_Q\psi_L\varphi_{Q^c} \right) +
  \text{h.c.} \qquad\text{and} \nonumber  \\
  \mathscr{L}_\ell & = h \left( \psi_{L}\psi_L\varphi_L +
    \psi_{Q}\psi_{Q}\varphi_Q + \psi_{Q^c}\psi_{Q^c}\varphi_{Q^c}
  \right) + \text{h.c.} \label{eq:YukawaLeptonFull}
\end{align}
On tree level, the minimal trinification model yields one active neutrino at the electro-weak scale and a $SU(2)_L$-singlet neutrino with a mass of a few eV.  Due to large radiative contributions these tree level results are corrected at one-loop level, giving rise to the radiative seesaw mechanism.  Then one light and two heavy neutrinos emerge, where the light neutrino is identified with the SM neutrino. However, as pointed out in Ref.~\cite{mt2006}, this mechanism is absent in a weak-scale supersymmetric extension of minimal trinification and other attractive approaches to the weak hierarchy problem.

\begin{figure}[htbp]
  \begin{center}
    \includegraphics[width=6cm]{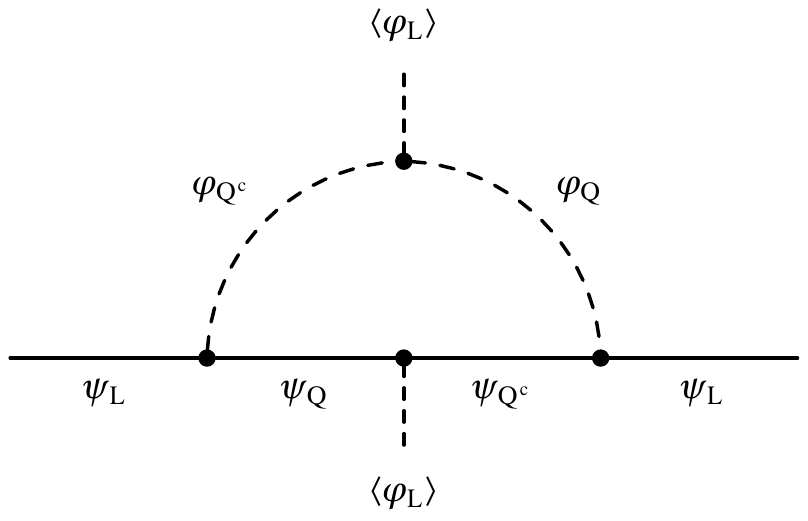}
    \hspace{0.5cm}
    \includegraphics[width=6.1cm]{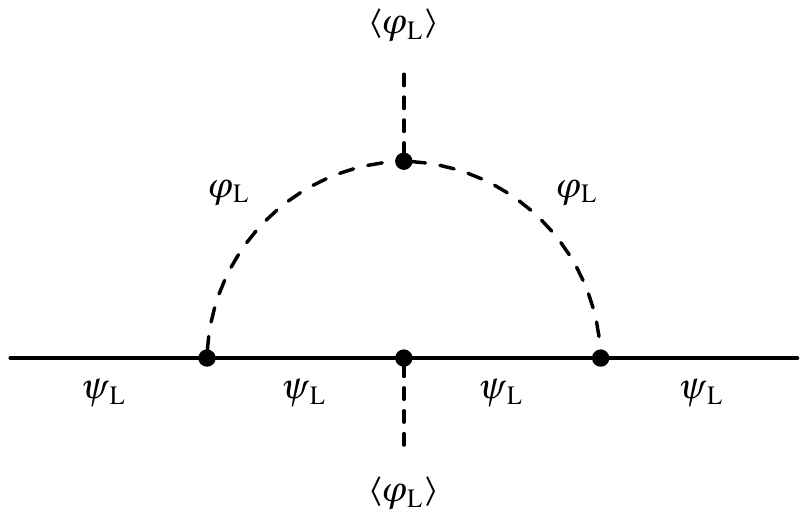}
  \end{center}
  \vspace{-0.5cm}
  \caption{One-loop contributions to neutrino masses via colored Higgs and fermion fields (left) and color-singlet Higgs and fermion fields (right). In the first case there is another diagram in which the Yukawa vertices are interchanged \cite{mt2006}.}
  \label{fig:OneLoop}
\end{figure}

With the given Lagrangian $\mathscr{L}=\mathscr{L}_q+\mathscr{L}_\ell+\mathscr{L}_h$, it is possible to construct the diagrams shown in Fig. \ref{fig:OneLoop}. 
While the left diagram contributes via colored Higgs and fermion fields involving the couplings $\psi_{Q^c}\psi_Q\varphi_L$, $\psi_L\psi_{Q^c}\varphi_Q$,$\psi_Q\psi_L\varphi_{Q^c}$ and $\varphi_{Q^c}\varphi_Q\varphi_L$, the right diagram uses the couplings $\psi_L\psi_L\varphi_L$ and $\varphi_L\varphi_L\varphi_L$ instead and contributes via color-singlet Higgs and fermion fields. 
Both diagrams are dominated by the fermion that acquires a unification-scale mass and are proportional to the mass of the involved Higgs fields in the loop. The diagrams are in the interaction basis.  In order to evaluate the contribution it is necessary to work in the mass eigenstate basis. 
After a straight-forward calculation, the dominant contribution of the left diagram, which is proportional to the fermion and the Higgs mass in the loop, gives the loop-factor
\begin{align}
    F_B 
      & = \frac{m_B}{(4\pi)^2} \frac{1}{2} \left(
      \frac{m_{B_{H\,1}}^2}{m_B^2-m_{B_{H\,1}}^2}
      \log\frac{m_{B_{H\,1}}^2}{m_B^2} -
      \frac{m_{B_{H\,2}}^2}{m_B^2-m_{B_{H\,2}}^2}
      \log\frac{m_{B_{H\,2}}^2}{m_B^2} \right ).
  \end{align}
  
The second diagram in Fig.\ref{fig:OneLoop} adds up in an analogous contribution where the quark $B$ is replaced by the lepton doublet $E$ and the Higgs fields $B_{H\,1,2}$ are replaced by $SU(2)_L$-doublet, color-singlet Higgs fields. The mass entries in Eq.~(\ref{eq:OneGenMassMatrix}), however, are proportional to both the loop-factor and the corresponding Yukawa couplings. As the latter are smaller for leptons, the leptonic contributions are smaller. Hence, the one-loop neutrino mass matrix for one generation in the ($\nu, N_1,N_2$) basis is given by
\begin{align} \label{eq:OneGenMassMatrix}
  M_N^{\mathrm{1-loop}} \simeq
  \begin{pmatrix}
    0 & -h_1u_2 & 0    \\
    -h_1 u_2 & s_{\alpha-\beta}c_\beta\, g^2 F_B &
    \left(s_{2\beta}s_\alpha-c_\alpha\right) g^2 F_B  \\
    0 & \left(s_{2\beta}s_\alpha-c_\alpha\right) g^2 F_B &
    c_{\alpha-\beta}s_\beta g^2 F_B
  \end{pmatrix},
\end{align}
where $\alpha$ and $\beta$ parameterize the mixing in the quark (to single out the light states $d^c$) and leptonic sector, respectively. This matrix has two eigenvalues of $\mathcal{O}(1)$ and one eigenvalue $\mathcal{O}(\epsilon^2)$, with $\epsilon\sim\frac{h_1 u_2}{g^2 F_B}$, 
\begin{align}
  m_{N_{1,2}}\sim g^2 F_B,\qquad\qquad m_\nu\sim\frac{h_1^2 u_2^2}{g^2
    F_B}
\end{align}

In order to obtain the correct values for the tau and top masses, we expect $h_1\simeq 0.1$, $g\simeq 1$ and \mbox{$u_{1,2}={\cal O}\left(10^2\,\text{GeV}\right)$} \cite{pdg2008}.

With a unification scale at approximately $10^{16}$\,GeV and $F_B\simeq\frac{1}{(4\pi)^2}M_U\approx6\cdot10^{13}$\,GeV, the mass of the light neutrino is of $\mathcal{O}(0.1-0.01\,\text{eV})$.  Thus, the two sterile neutrinos become heavy with masses at the unification scale, while the mass of the light standard model-like neutrino is
suppressed by a radiative seesaw \cite{witten1979}.

\section{The inverse seesaw mechanism}\label{sec:InverseSeesaw}
Extensions of the Standard Models introduce new physics which suppresses the one-loop contributions to ${\cal O}\left(1\,\text{TeV}\right)$.  In supersymmetry, sparticle loop contributions generate cancellations due to the non-renormalization of the superpotential in the limit of exact supersymmetry. Hence, the nonvanishing entries in the neutrino mass matrix are all of the same order.  The natural cut-off scale of such effects ranges from a few TeV in weak-scale supersymmetry up to several hundred TeV in scenarios with large extra dimensions.

We will show, however, that it is possible to generate light neutrino masses via radiative corrections in a TeV-scale extension of minimal trinification. With $\order{10^5-10^6\,\text{GeV}}$ loop contributions and certain assumptions, a modified inverse seesaw mechanism is capable of reproducing the observed neutral lepton mass spectrum with one light neutrino per family as well as the remaining standard model particle masses.

The original idea behind the inverse seesaw mechanism \cite{valle1986,bernabeu1987} is to introduce a new heavy $SU(2)\times U(1)$ singlet lepton $N$ with an effective mass term, $\mu N N$. The mass of the singlet lepton can be much smaller than the mass of the singlet lepton of the standard seesaw.  In this case the smallness 
of the neutrino mass is directly related to the smallness of $\mu$. For one generation, the mechanism is characterized by a mass matrix of the following shape, in the $(\nu,\nu^c,N)$ basis:
\begin{align} \label{eq:InverseSeesawMatrix}
  \mathcal{M} = 
  \begin{pmatrix}
    0 & m_D^T & 0  \\
    m_D & 0 & M^T  \\
    0 & M & \mu
  \end{pmatrix},
\end{align}
which yields a light, active neutrino with a mass  $\mathcal{O}(0.1\,\text{eV})$ \cite{valle2005},
\begin{align} \label{eq:InverseSeesawFormula}
  \left(\frac{m_\nu}{0.1\,\text{eV}}\right) & =
  \left(\frac{m_D}{100\,\text{GeV}}\right)^2
  \left(\frac{\mu}{1\,\text{keV}}\right)
  \left(\frac{M}{10^4\,\text{GeV}}\right)^{-2} .
\end{align} 

This matrix has the same structure as the mass matrix in minimal trinification, Eq.~(\ref{eq:OneGenMassMatrix}) in the basis $(\nu,N_1,N_2)$, with the loop factor $F_B\simeq\tfrac{M_U}{(4\pi)^2}$ and a unification scale of approximately $M_U\simeq2\times10^{16}$\,GeV. In order to achieve a light neutrino mass, we aim to match both matrices given in Eqs.~(\ref{eq:OneGenMassMatrix}) and (\ref{eq:InverseSeesawMatrix}). We start with the $(2,3)$ and the $(3,2)$ entry. Looking at the matrix in Eq. (\ref{eq:InverseSeesawMatrix}) and comparing it to Eq.~(\ref{eq:InverseSeesawFormula}), these entries are of $\order{10\,\text{TeV}}$.

In contrast to minimal trinification, the loop factor $F_B$ is reduced from $\mathcal{O}(10^{14}\,\text{GeV})$ to $\tfrac{1}{(4\pi)^2}M_\text{X}$, where the new mass scale $M_X$ is in the multi-TeV region.  The Yukawa coupling $h_1$ can be chosen such that $M_X\simeq \order{10^5\,\text{GeV}}$ and therefore $M\sim\order{1\,\text{TeV}}$ (see Eq.~(\ref{eq:InverseSeesawFormula})).

In addition, we need the ($2,2$)-entry to vanish and the ($3,3$)-entry to be $\order{1\,\text{keV}}$ in order to satisfy the inverse seesaw conditions. These requirements may be fulfilled by an appropriate choice of the remaining free parameters, the mixing angles $\alpha$ and $\beta$, 
\begin{align} 
  \label{eq:inverse-conditions}
  c_{\alpha-\beta}\; s_\beta = \order{10^{-9}}, \qquad
  s_{\alpha-\beta}\; c_\beta = 0 \ .
\end{align}
Let us choose $\alpha=\beta$, which may be explained by an appropriate flavor symmetry.  In this case, the ($3,3$)-entry is simply given by $s_\beta$ and requires nearly vanishing mixing in the lepton and the quark sector, $\alpha=\beta=\arcsin(10^{-9})\approx10^{-9}$ such that
\begin{align}
  M_N^{\mathrm{1-loop}}\sim 
  \begin{pmatrix}
    0 & 10\,\text{GeV} & 0    \\
    10\,\text{GeV} & 0 & 1\,\text{TeV}    \\
    0 & 1\,\text{TeV} & 1\,\text{keV}
  \end{pmatrix}.
\end{align} 
As in minimal trinification, two neutrinos are heavy while the third one is light,
\begin{align}
  m_{N_{1,2}}\simeq 1\,\text{TeV}, \qquad m_\nu=0.1\,\text{eV},
\end{align}
Alternatively, one might choose $\alpha\simeq 10^{-9}$ and  $\beta=\tfrac{\pi}{2}$, i.e., $\tan\beta\rightarrow\infty$. This scenario, however, is not feasible, as $\tan\beta=h_1v_1/h_2v_2$ and both $h_2$ and $v_2$ have to be different from zero in order to reproduce the electron masses and allow for breaking $SU(3)_L\times SU(3)_R$ to $SU(2)_L\times SU(2)_R\times U(1)$.  Thus this scenario is ruled out.

To ensure a valid symmetry breaking chain, we thus choose $\alpha=\beta=10^{-9}$.  These small mixing angles, however, are not consistent with the general setup of the trinified model, which can easily be seen as follows. In order to correctly describe the fermion masses and mixing angles, the Yukawa couplings may not be too small such that
\begin{align}
  \tan\alpha = \tan\beta \approx \frac{v_1}{v_2} =
  \mathcal{O}\left(10^{-9}\right) ,
\end{align} 
yielding $v_1\ll v_2$.  We can now take a look at the heavy fermion masses given and find \cite{mt2006}
\begin{align}
  m_B & = \sqrt{g_1^2v_1^2+g_2^2v_2^2} \approx g_2v_2 \simeq
  10^{16}\,\text{GeV}, \qquad m_E = \sqrt{h_1^2v_1^2+h_2^2v_2^2}
  \approx h_2v_2 \simeq 10^{16}\,\text{GeV}.
\end{align}

With the known masses for the top and the bottom quarks as well as the tau lepton, we can calculate the Yukawa couplings for this model. To determine an allowed parameter set, two conditions have to be fulfilled: One, the ratio of the bottom and the top quark masses; two, the squares of the weak VEVs at up $\left(247\,\text{GeV}\right)^2$,
\begin{align}
  \frac{m_b}{m_t} & = \frac{u_1}{u_2}\, s_\alpha = 0.0245 \,, \qquad
  \sqrt{u_1^2+u_2^2} = 247\,\text{GeV} .
\end{align}
Combining these two conditions yields the weak VEVs as functions of the quark mixing angle $\alpha$.
\begin{align} 
  u_1 & = \frac{247\,\text{GeV}}{\sqrt{1+\left(\frac{m_t}{m_b}\,
        \sin\alpha\right)^2}}, \qquad u_2 =
  \frac{247\,\text{GeV}}{\sqrt{1+\left(\frac{m_b}{m_t}\,
        \csc\alpha\right)^2}} \ .
  \label{eq:ModSeesawVEVs}
\end{align}
The couplings $g_1$ and $h_1$ are then given by 
\begin{align}
  g_1 & = \frac{m_t}{u_2}, \qquad h_1 = \frac{m_\tau}{u_1
    \sin\,\beta} \ .
  \label{eq:ModSeesawYukawaCouplings}
\end{align}
Since $\alpha=\beta=10^{-9}$, we obtain 
\begin{align}
    u_1 & \simeq 247\,\text{GeV}, \quad u_2 \simeq
    10^{-5}\,\text{GeV}, \quad g_1 \simeq 1.7\times 10^7 \quad
    \text{and} \quad h_1 \simeq 6.9 \times 10^6 \ .
  \end{align}
These large values for the Yukawa coupling are an obvious problem. So we can either choose reasonable values for the Yukawa couplings or for the vacuum expectation values, but not for both.

\section{A Modified Inverse Seesaw}
As the two conditions for the inverse seesaw mechanism cannot be fulfilled simultaneously within our model, we investigate whether it is possible to relax either of them. We will therefore have another look at the neutrino mass matrix, Eq.~(\ref{eq:OneGenMassMatrix}). The conditions listed in Eq.~(\ref{eq:inverse-conditions}) stem from the $(2,2)$- and $(3,3)$-entries; however, it is obvious that the latter one accounts for the inverse seesaw mechanism as it introduces the large hierarchy among the entries.  Then the $(2,2)$-entry simply has to be chosen such that it does not spoil the mechanism. Of course, relaxing the condition on the $(2,2)$-entry, i.e., allowing for larger mixing angles, has an impact on the $(2,3)$-entry. Its value is bound from below by the masses of the heavy neutral leptons of $90.3$\,GeV \cite{pdg2008}, which corresponds to a lower bound on the $(2,3)$-entry of about $330$\,GeV.

Let us therefore consider the mass matrix,
\begin{align}\label{eq:ModSeeSawMassMatrix}
  M_N^{\mathrm{1-loop}} =
  \begin{pmatrix}
    0 & m_D & 0    \\
    m_D & \tilde{M} & M    \\
    0 & M & \mu
  \end{pmatrix}
  \sim
  \begin{pmatrix}
    0 & 10\,\text{GeV} & 0    \\
    10\,\text{GeV} & 0-1\,\text{TeV} & 0.33-1\,\text{TeV}    \\
    0 & 0.33-1\,\text{TeV} & 1\,\text{keV}
  \end{pmatrix}
  .
\end{align}
In fact, this kind of matrix was already considered in the context of inverse type-III seesaw models \cite{ibanez2009}. It is straight-forward to show that two eigenvalues are $\order{1\,\text{TeV}}$, while the third one is indeed given by $\order{(0.1-1)\,\text{eV}}$,
\begin{align}
  \lambda_h \approx \frac{1}{2} \left(\tilde{M} +
    \mu\pm\sqrt{4M^2+4m_D^2+\tilde{M}^2+\mu^2-2\tilde{M}\mu} \right) ,
  \qquad \lambda_\ell \approx \frac{m_D^2\mu}{M^2} \ .
\end{align}

We want to show now, that the doublet neutrino $\nu$ becomes light while the singlet neutrinos $N_{1,2}$ remain massive enough to evade the experimental bounds. An analytical determination of the eigenspace of the mass matrix is not possible due to the approximation used during the calculation of the eigenvalues. Instead we constrain ourselves to a numerical analysis. The eigenvectors of $M_N$ (Eq.\ref{eq:ModSeeSawMassMatrix}) are\footnote{For the ($2,2$) entry  $\tilde{M}=M=1$\,TeV has been chosen, as the exact value of $\tilde{M}$ has no influence on the mass eigenstates obtained. However with a vanishing $\tilde{M}$ we get exactly maximal mixing between the two heavy sterile states.}
\begin{align}
  w_1&\simeq(-1,\; -10^{-11},\; 0.01)\nonumber\\
  w_2&\simeq(0.009,\; -0.5,\; 0.9)\\
  w_3&\simeq(0.005,\; 0.9,\; 0.5).\nonumber
\end{align}

So the rotation matrix $S$ to diagonalize the mass matrix $M_N^\text{diag}=S^TMS$ can be constructed using the eigenvectors as columns.
\begin{align}
  S\approx
  \begin{pmatrix}
    -1 & 0.009 & 0.005\\
    -10^{-11} & -0.5 & 0.9\\
    0.01 & 0.9 & 0.5
  \end{pmatrix}
\end{align}

\begin{figure}[ht]
  \begin{center}
    \includegraphics[width=10cm]{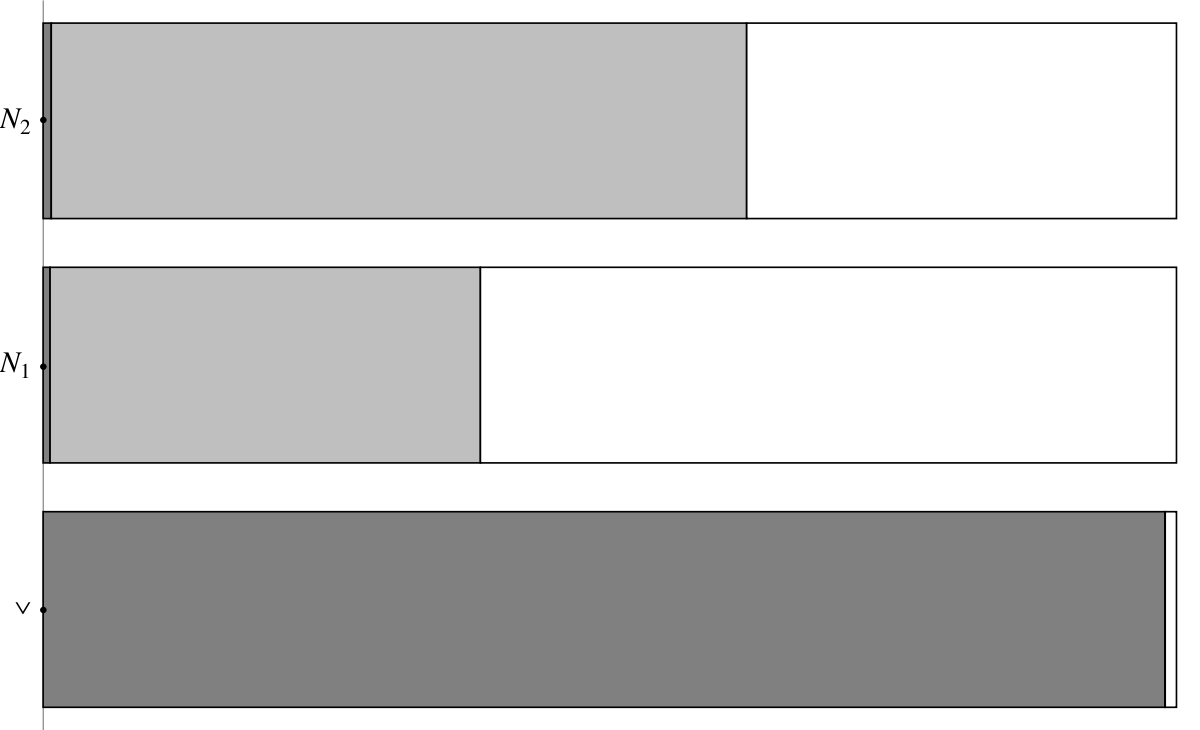}
    \caption{Neutrino mixing in a modified inverse seesaw model. The three different mass eigenstates are given by three different gray tones. While for the lightest neutrino $\nu$ mixing is almost absent (as required by unitarity of the light neutrino PMNS mixing matrix), the two heavy neutrinos $N_1$ and $N_2$ show up a nearly maximal mixing in mass eigenstates.}
    \label{fig:ModSeeSawNeutrinoMixing}
  \end{center}
\end{figure}

\begin{figure}[htbp]
 \includegraphics[width=7.12cm]{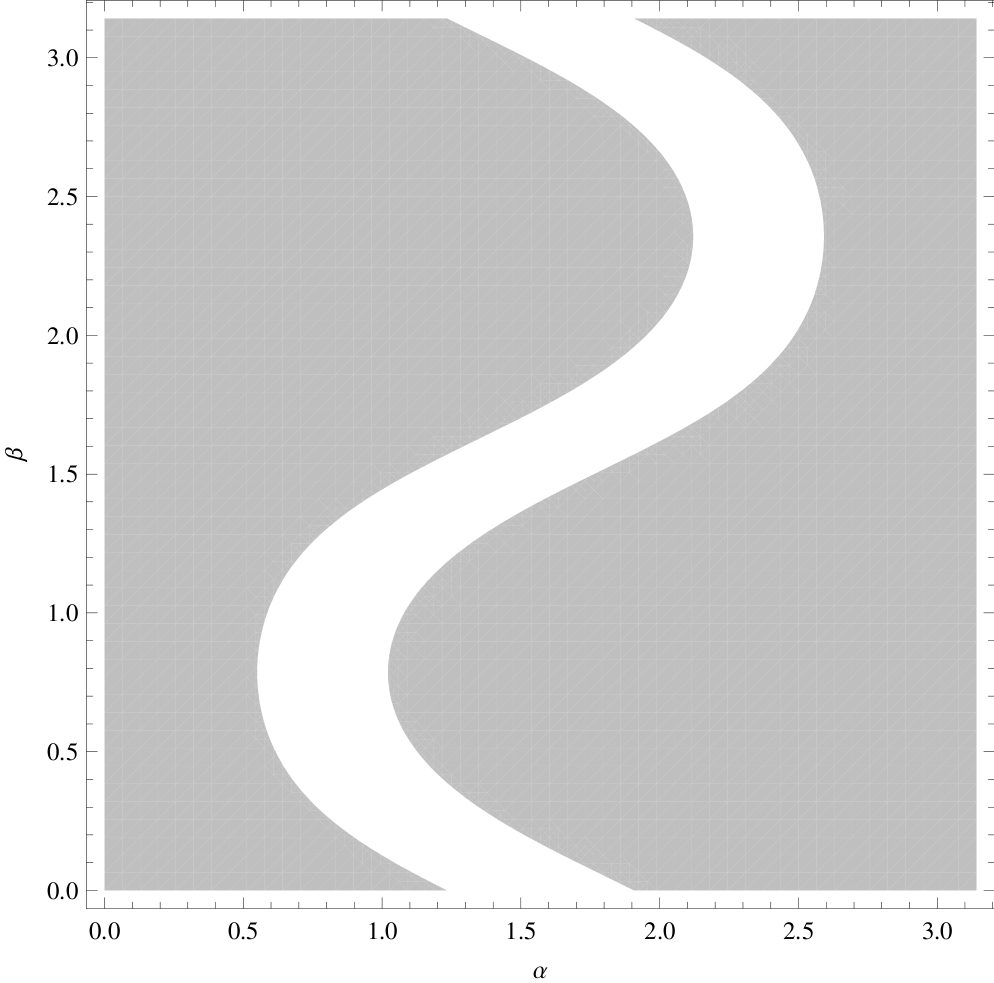}
  \includegraphics[width=7cm]{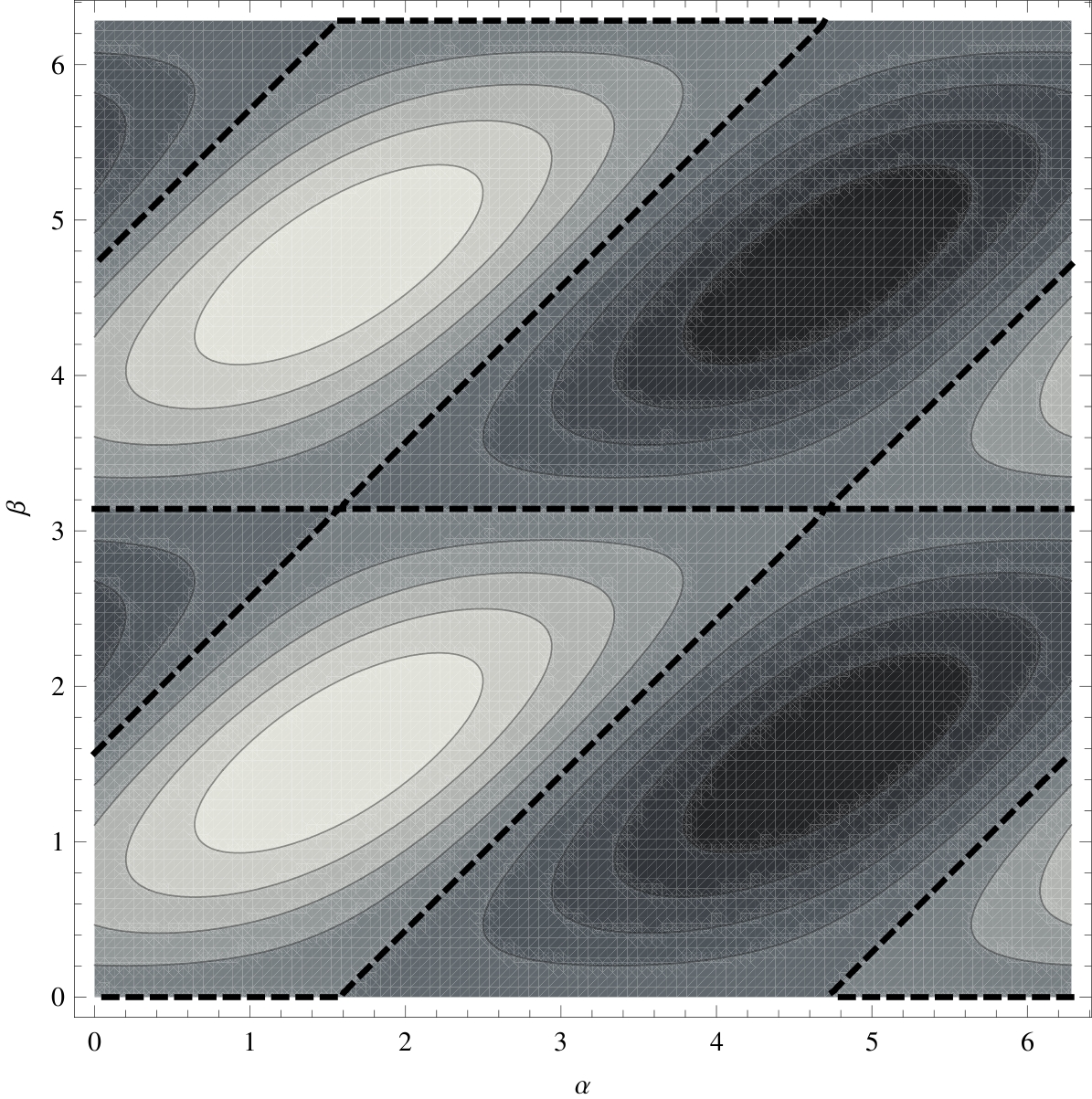}
  \caption{The left plot shows the allowed range of values for $\alpha$ and $\beta$ to fulfill the condition $s_{2\beta}s_\alpha-c_\alpha\geq 0.33$ from the bound on the $(2,3)$ and $(3,2)$ elements. The white band marks the forbidden area. On the second plot the contour describes the value of the ($3,3$) entry, where the dashed lines indicate the allowed values for $\alpha$ and $\beta$ to get an ($3,3$) entry equal $10^{-9}$.}                                                   \label{fig:RegionPlotAngleCondition}
\end{figure}

\begin{figure}[htbp]
 \begin{center}
  \includegraphics[width=7cm]{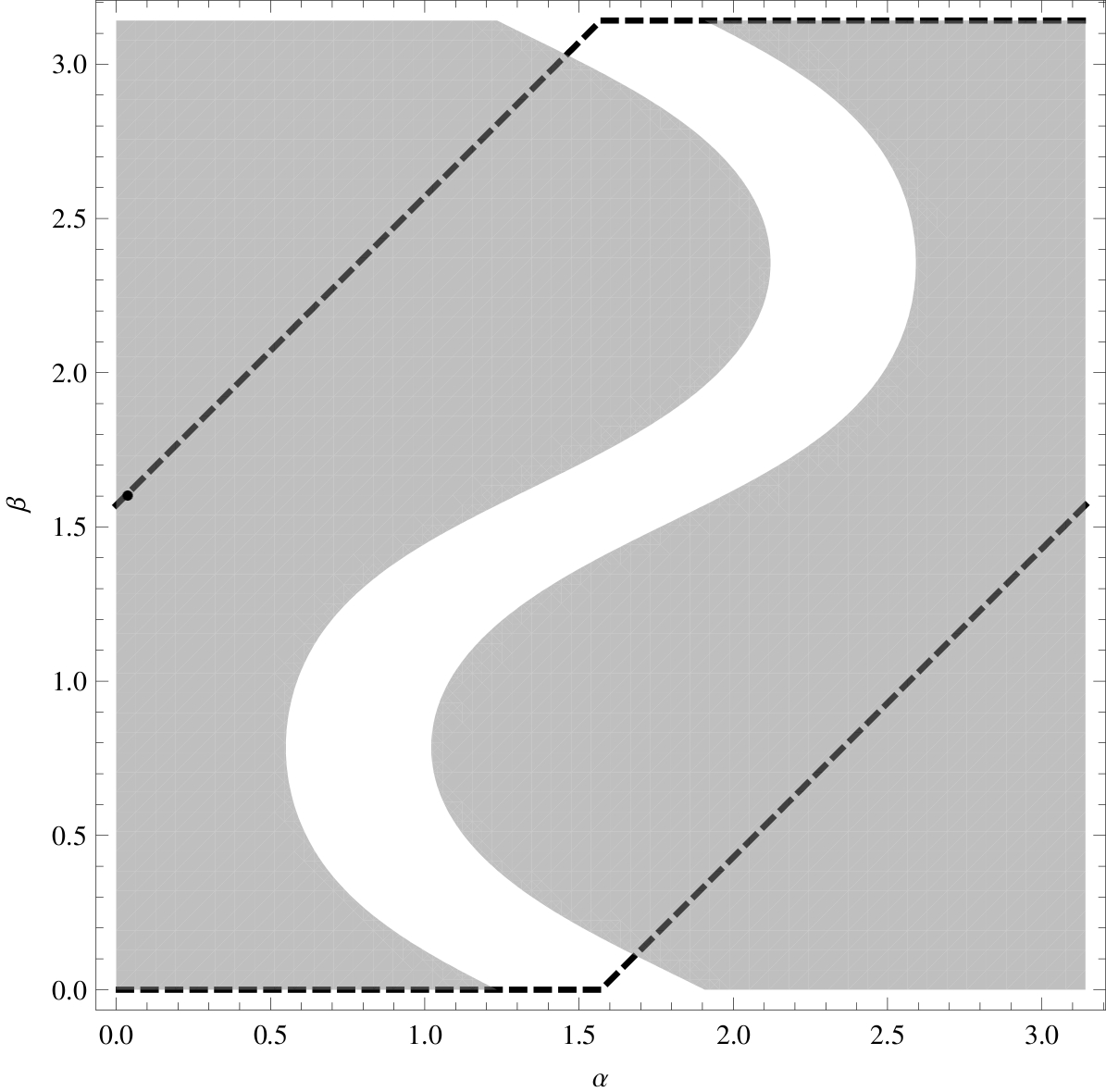}
   \caption{Overlay of the two plots in Fig. \ref{fig:RegionPlotAngleCondition}. The white band marks the forbidden area, with respect to the experimental bounds of new heavy neutral leptons. 	The dashed lines indicate the allowed values for $\alpha$ and $\beta$ to get an ($3,3$) entry equal $10^{-9}$.}
    \label{fig:OverlayPlotAngleCondition}
 \end{center}
\end{figure}

This yields a minimal mixing of the lightest neutrino $\nu$ and a nearly maximal mixing between $N_1$ and $N_2$ as shown in Fig.\ref{fig:ModSeeSawNeutrinoMixing}. 
Now we want to determine the conditions on $\alpha$ and $\beta$ to fulfill the scheme introduced in Eq. (\ref{eq:ModSeeSawMassMatrix}). At first there is again $c_{\alpha-\beta}s_\beta = \pm10^{-9}$, which is crucial for the light neutrino mass. Additionally the second condition $s_{2\beta}s_\alpha-c_\alpha\geq 0.33$ has to be fulfilled\footnote{As mentioned in the last subsection, the ($3,2$)- and the ($2,3$) entry must not be smaller than $330$\,GeV. See Eq. (\ref{eq:ModSeeSawMassMatrix}).}. Fig. \ref{fig:RegionPlotAngleCondition} shows the allowed range of values. The overlay of both plots in Fig. \ref{fig:OverlayPlotAngleCondition} shows, that large portions of the parameter space of each individual angle $\alpha$ and $\beta$ are allowed, while the values of the angles are strongly correlated. The dashed line shows this correlation between $\alpha$ and $\beta$.

We discard the case where $\beta$ is nearby $0$ or $\pi/2$ since this choice does not allow for symmetry breaking to the standard model, as was discussed in section \ref{sec:InverseSeesaw}. Instead we concentrate on the other region where $\beta$ is given by 
\begin{align}\label{eq:AlphaBeta1}
  \beta=\arccos\left(-\frac{1}{2}\sqrt{3-\cos\,2\alpha-4\times
      10^{-9}\sin\,\alpha-2\sqrt{\cos^2\alpha\left(\cos^2\alpha+4\times
          10^{-9}\left(\sin\,\alpha-10^{-9}\right)\right)}}\right)
\end{align}
for $0\leq\alpha<\pi/2$, and 
\begin{align}\label{eq:AlphaBeta2}
  	\beta=\arccos\left(\frac{1}{2}\sqrt{3-\cos\,2\alpha-4\times
      10^{-9}\sin\,\alpha-2\sqrt{\cos^2\alpha\left(\cos^2\alpha+4\times
          10^{-9}\left(\sin\,\alpha-10^{-9}\right)\right)}}\right)
\end{align}
for $\pi/2<\alpha\leq\pi$.

We determine the parameter spectrum for the fermion and gauge boson masses as before. Using the correlation of $\alpha$ and $\beta$ given in Eqs.~(\ref{eq:AlphaBeta1}) and (\ref{eq:AlphaBeta2}) we are able to calculate the VEVs and the Yukawa couplings with respect to a specific value of the quark mixing angle $\alpha$. 
\begin{figure}[htbp]
  \centering
  \begin{minipage}[t]{7cm}
    \includegraphics[width=7cm]{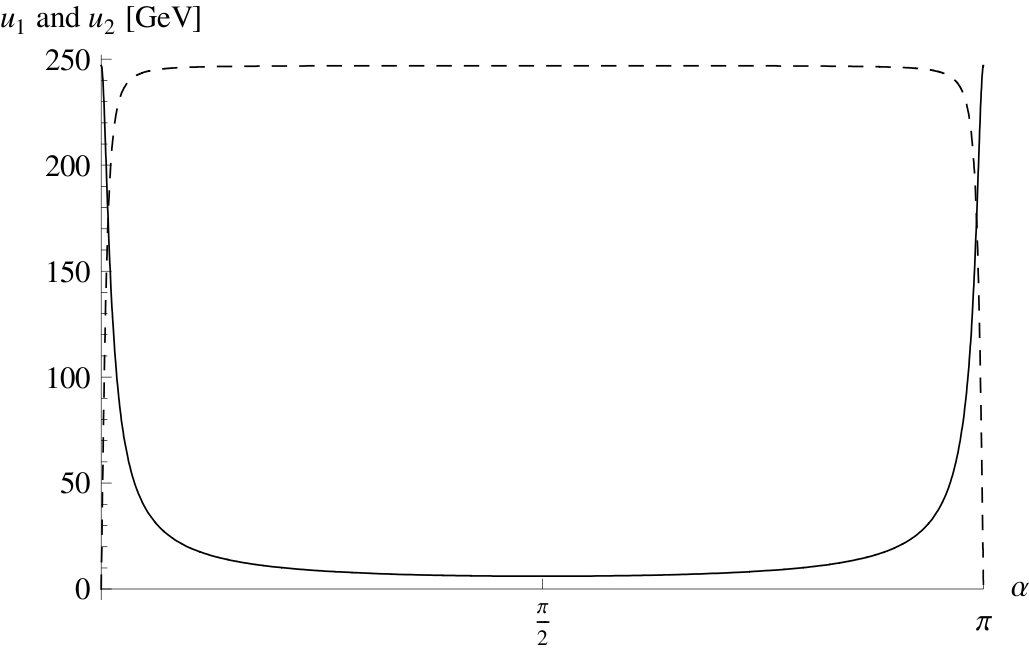}
    \caption{The vacuum expectation values $u_1$ and $u_2$ as a function of the quark mixing angle $\alpha$. The solid line corresponds to $u_1$, the dashed one to $u_2$}	   \label{fig:VEVRatioThirdGeneration}
  \end{minipage}\hfill
  \begin{minipage}[t]{7cm}
    \includegraphics[width=7cm]{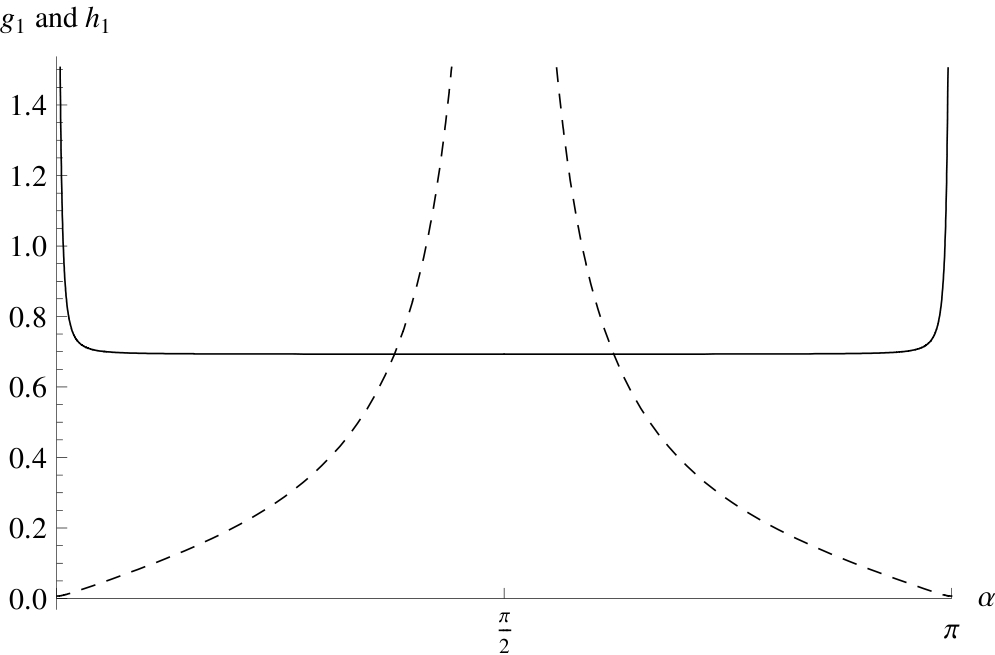}
    \caption{Both Yukawa couplings, $g_1$ for the quark sector and $h_1$ for the lepton sector are plotted against the quark mixing angle $\alpha$. The solid line corresponds to $g_1$, the dashed one to $h_1$.}
    \label{fig:YukawaCouplingsThirdGeneration}
  \end{minipage}
\end{figure}
Figs. \ref{fig:VEVRatioThirdGeneration} and \ref{fig:YukawaCouplingsThirdGeneration} show the weak-scale vacuum expectation values and the Yukawa couplings $g_1$ and $h_1$ as a function of $\alpha$, respectively. We notice that part of these parameters become divergent for the mixing angles $\alpha=0$, $\pi/2$ and $\pi$. These regions, however, are already excluded, as discussed above, so the parameters fulfill all provided conditions in the remaining regions.

\begin{figure}[htbp]
  \begin{center}
    \includegraphics[width=7cm]{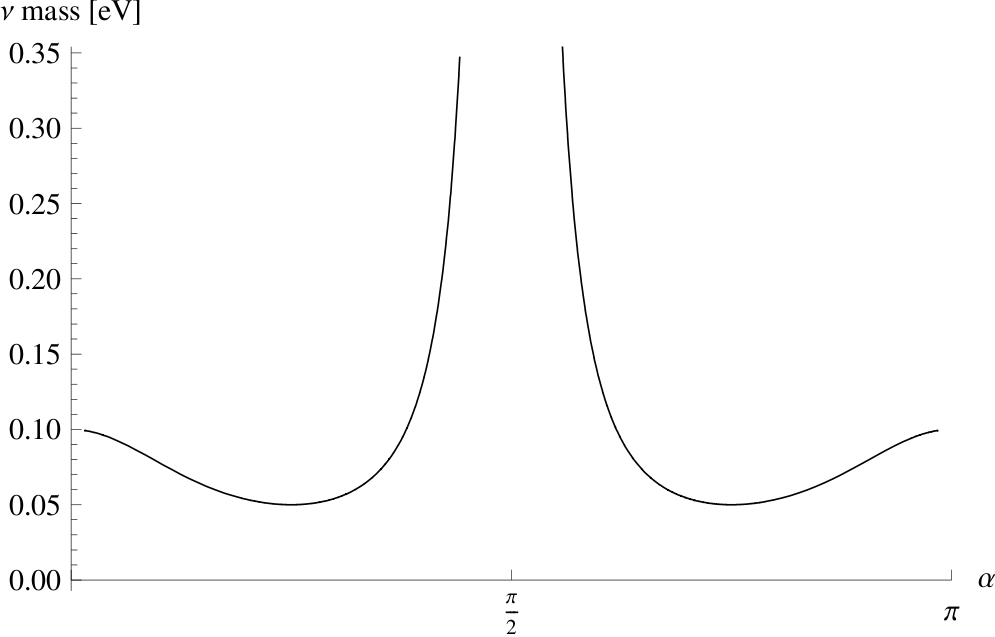}
    \caption{The mass of the lightest neutrino as a function of the quark mixing angle $\alpha$. Naturally this behavior depends also on the restrictions for the $(3,3)$ matrix entry. The mass function diverges around $\pi/2$ as well as for $\alpha=0$ and $\pi$, which is not clearly visible in this plot.}
   \label{fig:LightNeutrinoMass}
  \end{center}
\end{figure}

\begin{figure}[htbp]
  \includegraphics[width=7cm]{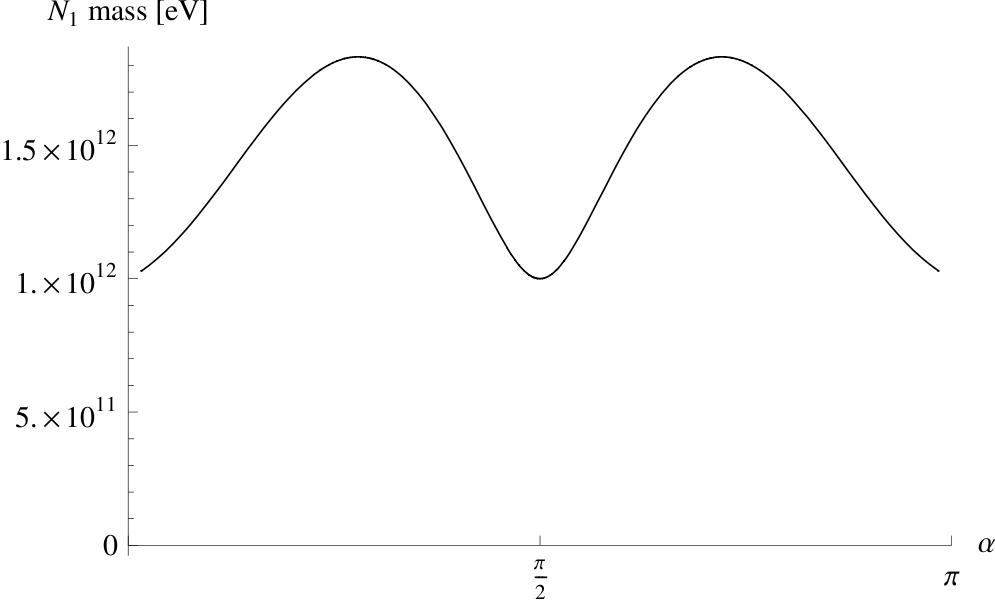}
  \includegraphics[width=7cm]{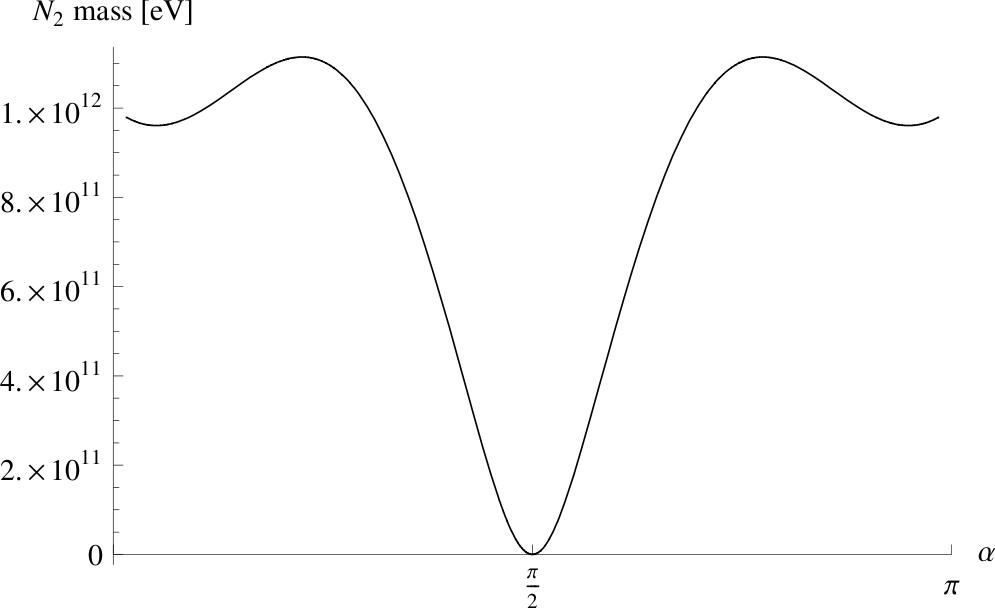}
  \caption{The mass of $N_1$ and $N_2$ as a function of $\alpha$. Again the masses depend on the restrictions for the $(2,3)$, $(3,2)$, and $(3,3)$ matrix entries which have been chosen to fulfill the experimental bounds from searches for new heavy neutral leptons.}
  \label{fig:HeavyNeutrinoMasses}
\end{figure}

All values except for the regions around $0$, $\pi/2$, and $\pi$ imply a neutrino mass around $0.1\,$eV for the $SU(2)$-doublet neutrino. By varying the condition $c_{\alpha-\beta}s_\beta\simeq\order{10^{-9}}$ it is furthermore possible to achieve even smaller SM neutrino masses down to $\order{10^{-4}\,\text{eV}}$ without changing the masses of the new heavy neutrinos.  Figs. \ref{fig:LightNeutrinoMass} and \ref{fig:HeavyNeutrinoMasses} show the light and heavy neutrino masses as functions of $\alpha$, respectively. The masses of the heavy neutrinos $N_1$ and $N_2$ are smaller around the well known points $0$, $\pi/2$, and $\pi$. In particular $N_2$ becomes as light as a few hundred MeV for $\alpha=\pi/2$.

It is evident that the strong correlation between the quark and lepton mixing angles $\alpha$ and $\beta$ requires a considerable amount of fine-tuning. This could be obtained by an appropriate flavor symmetry which would protect the relation against renormalization effects as well. Moreover, this model has to be compared to minimal trinification where gauge-coupling unification results if five Higgs doublets are at the weak scale, without supersymmetry, as well as to SUSY models based on $SU(5)$ or $SO(10)$ which require intermediate scales, additional Higgs fields, or higher-dimensional operators to correctly describe the fermion masses and mixing angles.

\section{Proton Decay and Lifetime}
In minimal trinification, proton decay is mediated at tree level, through Yukawa interactions involving the colored Higgs fields $\varphi_Q$, $\varphi_{Q^c}$. Because of the same baryon number assignment in both the quark and the lepton Yukawa Lagrangian a $\varphi_L$ mediated proton decay is forbidden. In general, there are two distinct types of operators, namely those involving only left or right-handed fields, {\slshape LLLL} and {\slshape RRRR}, and the mixed operators, {\slshape LLRR} and {\slshape RRLL}. The mixed operators read \cite{mt2006}
\begin{align}\label{eq:OpMixed}
    \mathscr{L}_\text{mixed} & \propto \left( g^{\ast ij} h^{mn}\, Q_m
    Q_n e^{c\ast}_i u^{c\ast}_j + \left( g\,\hat s_\beta \right)^{ij}
    \left( -\hat s_\alpha^\top\,h \right)^{\ast mn}\, d^{c\ast}_m
    u^{c\ast}_n Q_i L_k \right) + \text{h.c.} ,
\end{align}
where we explicitly display the generation indices.  They are of mass-dimension six; a unification scale of $\mathcal{O}\left(10^{16}\,\text{GeV}\right)$, as is the case in the supersymmetric model, sufficiently suppresses the corresponding decay rate. The {\slshape LLLL} and {\slshape RRRR} operators, however, can be of mass-dimension five,
\begin{align}\label{eq:OpDim5}
    \mathscr{L}_\text{dim5} & \propto \left( \left(g\,\hat s_\beta
    \right)^{ij} h^{mn}\, Q_m Q_n Q_i L_j + g^{ij} \left( -\hat
    s_\alpha^\top\,h \right)^{mn}\, d^c_m u^c_n e^c_i u^c_j \right)
    \vphantom{\frac12} + \text{h.c.}
\end{align}
In the presence of supersymmetry, they stem from F terms.  When the sfermions are integrated out, they give rise to effective four-fermion operators of dimension six. Thus the operators are suppressed by \mbox{$\left(m_s M_\text{U}\right)^2$} instead of $M_\text{U}^4$.  

It is beyond the purpose of this paper to calculate the lifetime of the various decay channels. It is well known that in ordinary SUSY GUTs, the decay rate is naturally consistent with the experimental limit if the sfermion masses, $m_s$, are above a few hundred TeV. (The PeV-scale as the \enquote{best place for supersymmetry} was discussed in \cite{wells2004}.) Moreover it is remarkable that the operators in Eqs. (\ref{eq:OpMixed}) and (\ref{eq:OpDim5}) are naturally suppressed for those choices of the mixing angles $\alpha$ and $\beta$ which also predict small neutrino masses.

\section{Conclusion}
Neutrino masses in TeV-scale extensions of Minimal Trinification have been a challenge as they either request higher-dimensional operators or large Higgs representations. In this paper, we have presented a new approach for light neutrino masses: a radiatively generated inverse seesaw mechanism with loop contributions from the PeV-scale. We have discussed how the basic mechanism can be implemented into minimal trinification and demonstrated that a realistic pattern of fermion masses can be obtained. If we compare the specific constraints which are set on the model parameters with the so-far discussed scenarios in the literature, we note that both higher-dimensional operators and large Higgs representations introduce additional sets of free parameters, as does minimal trinfication which requires several Higgs fields. Models based on other gauge groups (like $SU(5)$ or $SO(10)$) share these issues as well. Hence, the model discussed in this paper is a viable and as attractive scenario: it is able to generate neutrino masses in the $0.1$\,eV region if the mixing patterns in the quark and lepton sectors are correlated. This correlation might be explained by an appropriate flavor symmetry.

The existence of new gauge singlet neutrino states with large Yukawa couplings around the TeV scale has potentially interesting consequences for phenomenology. While lepton number violation in the inverse seesaw is supressed by the smallness of the parameter $\mu$ in the inverse seesaw mass matrix, large lepton flavor violating effects can arise. For example, enhanced rates for $\mu \to e \gamma$ as well as $\mu$-e conversion in nuclei with respect to seesaw-I model expectations both with and without supersymmetry have been found in \cite{deppisch2005, valle2005}.

Another exciting possibility would be the direct production of the gauge singlet neutrinos at the LHC. As lepton number is almost conserved, this scenario resembles the production of heavy Dirac neutrinos at the LHC discussed in \cite{delAguila2008}. In this paper a 5~$\sigma$ discovery reach for heavy neutrino masses up to 100~GeV was advocated with 30~fb$^{-1}$. While for larger masses the production cross section would decrease, new decay channels open up once the heavy neutrino mass exceeding the Higgs mass, which would require a detailed simulation.

\bibliography{literature}
\end{document}